# Numerical simulations of heat transfer in plane channel flow

Najla EL GHARBI [1, 3, a], Rafik ABSI [2, b] and Ahmed BENZAOUI [3, c]

[1] Renewable Energy Development Center, BP 62 Bouzareah 16340 Algiers, Algeria

[2] EBI, Inst. Polytech. St-Louis, 32 Boulevard du Port, 95094 Cergy-Pontoise, France

[3] University of Sciences and Technology Houari Boumediene, Algiers, Algeria

[a]n.elgharbi@cder.dz ; [b]r.absi@ebi-edu.com ; [c]abenzaoui@gmail.com

**Keywords:** turbulent flows, heat transfer, forced convection, low Reynolds number model.

**Abstract.** Reynolds-averaged Navier–Stokes (RANS) turbulence models (such as k-ε models) are still widely used for engineering applications because of their relatively simplicity and robustness. In fully developed plane channel flow (i.e. the flow between two infinitely large plates), even if available models and near-wall treatments provide adequate mean flow velocities, they fail to predict suitable turbulent kinetic energy "TKE" profiles near walls. TKE is involved in determination of eddy viscosity/diffusivity and could therefore provide inaccurate concentrations and temperatures. In order to improve TKE a User Define Function "UDF" based on an analytical profile for TKE was developed and implemented in Fluent. Mean streamwise velocity and turbulent kinetic energy "TKE" profiles were compared to DNS data for friction Reynolds number $Re_\tau = 150$. Simulation results for TKE show accurate profiles. Simulation results for horizontal heated channel flows obtained with Fluent are presented. Numerical results are validated by DNS data for $Re_\tau = 150$.

## Introduction

Turbulent flow with heat transfer mechanism is of great importance from both scientific and engineering field because it occurs frequently in many industrial applications, such as heat exchangers, gas turbine cooling systems, nuclear reactors ... To simplify the geometry and to understand the mechanism of transport, the flow in a channel, has been studied extensively from both experimental and numerical approaches.

During last years, the performance of turbulence and heat transfer models in predicting the profiles of velocity and temperature in industrial flows has become increasingly important, especially with recent progress in the CFD "Computational Fluid Dynamics" field. Consequently, several researchers preferred the use of the numerical approaches because they are simpler and less expensive than the experimental approaches. The direct numerical simulation (DNS) is a robust method where several researchers developed their investigations. Kim and Moin [1] studied the heat transfer in two dimensional forced convection channel flow for different Prandtl numbers. Kasagi et al. [2] revisited the problems employing a constant time-averaged heat flux boundary condition on the walls, for a mild Reynolds number of 4580. Kawamura et al. [3] analyzed the effects due to the change of Reynolds and Prandtl numbers on the heat turbulent transport; they also compared various boundary conditions for velocity and temperature. Tiselj et al. [5] took into account the fluctuation of temperature close to walls whereas in previous studies, the flow of heat was taken as constant (ideal case).

Turbulence models are widely used for simulating complex heat transfer and flow phenomena in many engineering applications because of their simplicity and effectiveness. The most popular is the standard k-ε model (High-Reynolds model, High-Re) proposed by Launder and Spalding [6].

However, the disadvantage of the standard k-ε model with standard wall functions is the inability to predict accurate near wall flow characteristics. To solve the near-wall effect, a number of Low-Reynolds-Number models (LRN model) have been developed, such as special treatment in this region.

The first LRN *k-ε* model was developed by Jones and Launder [7] and subsequently modified by many researchers. To get a better understanding of the near wall effect, many LRN models were proposed by introducing damping functions and other additional terms. Most of LRN models were developed based on the High-Re *k-ε* model (Lam and Bremhorst [8], Chien [9], Abe et al., [10], Chang et al., [11]). Some other new LRN two-equation models have been proposed as alternatives to the LRN *k-ε* models, e.g., the standard k–ω model and its LRN variant by Wilcox [12]. Menter [13] proposed SST *k-ω* model to resolve the free-stream dependency by blending the standard Wilcox model and the standard *k-ε* model.

The aim of our study is to improve the prediction of the flow in the near wall region. A near-wall function for TKE will be implemented in Fluent through a UDF. Three turbulent model will be tested with DNS data: *k-ε* standard with enhanced wall treatment [14], *SST k-ω* and *k-ε* models with low Reynolds (model of Chang et al.).

## Simulation procedures

**Test case.** A fully developed forced convection flow in a parallel channel is numerically simulated. The steady turbulent Navier-Stokes and energy equations are numerically solved together with the continuity equation using the finite volume method. Physical properties are considered as constants and evaluated for air at the inlet temperature $T_0 = 20°C$ with a density $\rho = 1.205$ kg/m3, molecular dynamic viscosity $\mu = 1.82 \times 10^{-5}$ Kg/ms, specific heat $Cp = 1005$ J/Kg°C, and thermal conductivity $\lambda = 0.0258$ W/m°C.

Flow at the inlet section of the channel is considered to be isothermal ($T_0=20°C$), with a uniform streamwise velocity component ($u$). The other velocity component ($v$) is set to be equal to zero at that inlet section. Periodic boundary conditions were used in the streamwise and spanwise directions and no-slip boundary conditions were imposed at the solid walls.

We considerate two case: (1) the walls is supposed isothermal (2) the walls were with uniform heat flux ($q_w$=500 W/m2).

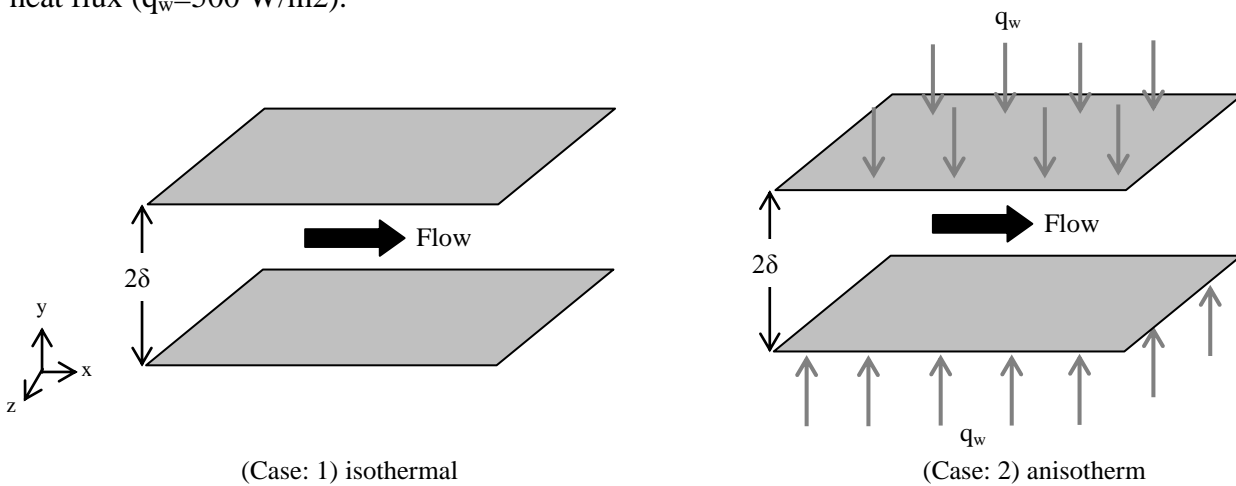

(Case: 1) isothermal          (Case: 2) anisotherm

Figure 1. Sketch of channel flows

**Tools.** A commercial code Fluent 6.3 was used, with a fine mesh near the wall. The number of mesh is (41 ×100). A second order scheme is used to discretize the diffusion terms. The solution is obtained in primitive variable P-V, where the coupled pressure and velocity equations are solved by the simple algorithm. The convergence criterion was based on a maximum error less than a prescribed value, taken equal to $10^{-8}$ for energy, and $10^{-6}$ for the other equations.

## Results and discussion

To study the flow near the wall, we tested three turbulent model compared to DNS of Iwamato et al.and DNS of Kasagi et *al.*: $k$-$\varepsilon$ standard with enhanced wall treatment [14], *SST k-ω* and $k$-$\varepsilon$ models with low Reynolds (model of Chang et al.). These models are able to predict the flow near a wall.

**Isothermal case.** According to Figure 3, the $k$-$\varepsilon$ model with enhanced wall Standard Treatment and the *SST k-ω* model provides good velocity profile throughout the wall on the other side the low Reynolds model gives good results except in the region $y^+ > 30$. The turbulent kinetic energy is well provided by $k$-$\varepsilon$ standard with enhanced wall Treatment in the region $y^+ \leq 30$, and by *SST k-ω* and low Reynolds model in the region $y^+ > 50$, compared to the DNS. So to improve the velocity and the turbulent kinetic energy profiles in the region $y^+ \leq 30$ we will implemente a function [15] Eq.1 in Fluent through a UDF [16]:

$$k^+ = B(y^+)^2 e^{\left[-\frac{y^+}{A}\right]} \qquad (1)$$

Where *A* and *B* are two parameters, $A=8$ and *B* is dependent on friction Reynolds number $Re_\tau$ (defined by friction velocity $u_\tau$, kinematic viscosity $\nu$ and the channel half-width δ) and given by $B = C_{B1} \ln(Re_\tau) + C_{B2}$ [17].

Table 1. Values of coefficient $B\ (Re_\tau)$ obtained from Eq.1 and DNS

| $Re_\tau$ | 110  | 150 | 300   | 400  | 590   | 642  |
|-----------|------|-----|-------|------|-------|------|
| $B$       | 0,09 | 0,1 | 0,115 | 0,12 | 0,128 | 0,13 |

Table 1 gives values of $B\ (Re_\tau)$ obtained from Eq.1 implemented in Fluent with DNS data [18, 2]. Our calibration of *B* with Fluent gives $C_{B1} = 0.0219$ and $C_{B2} = -0.0113$ (Fig.2).

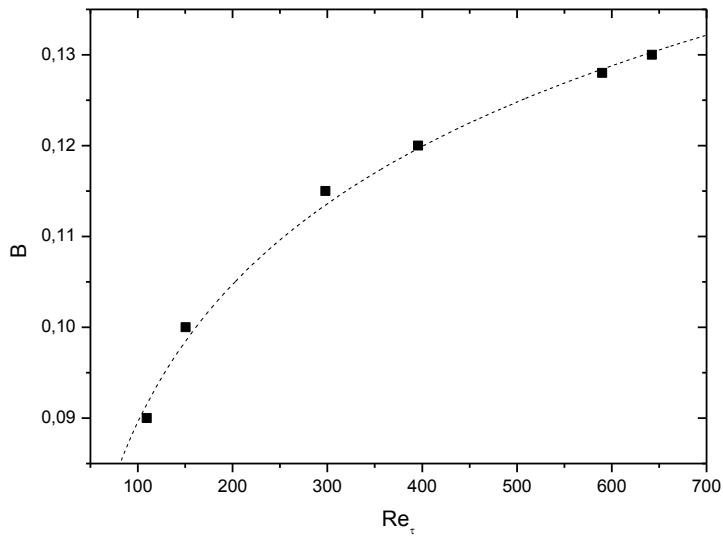

Figure 2: Dependency of the coefficient *B* on the Reynolds number $Re_\tau$.
values obtained from DNS data and Fluent, curve, proposed function

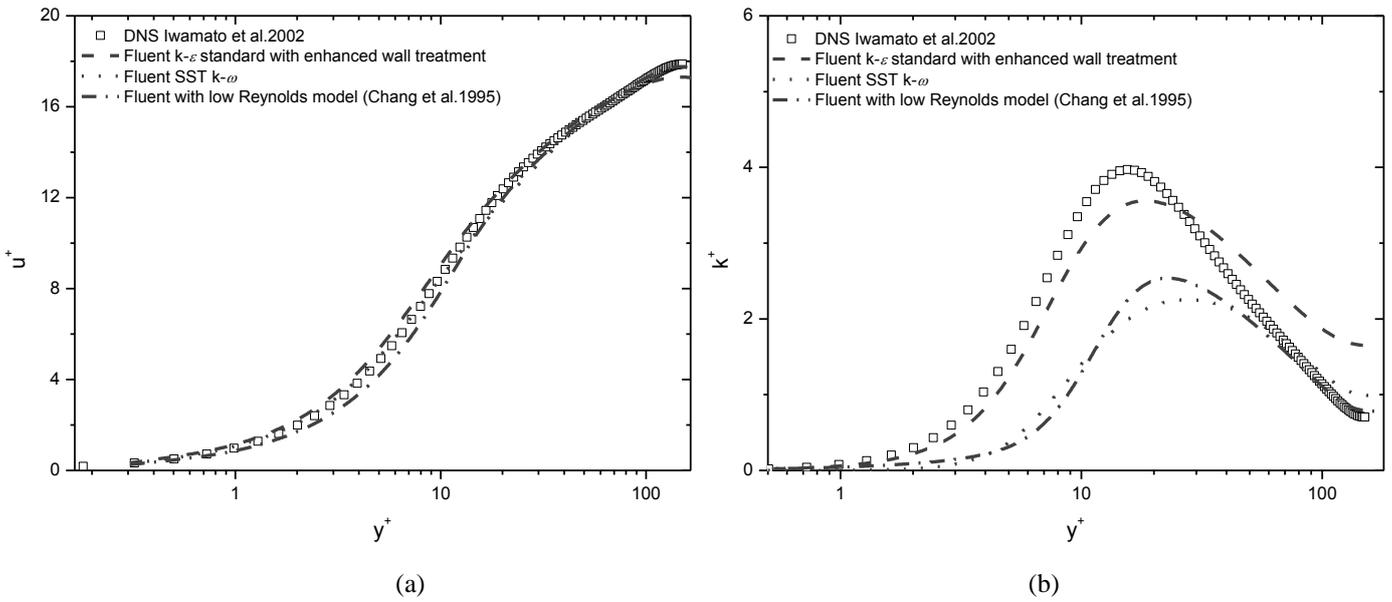

Figure 3. Comparison of (a) velocity (b) turbulent kinetic energy profiles models for $Re_\tau = 150$

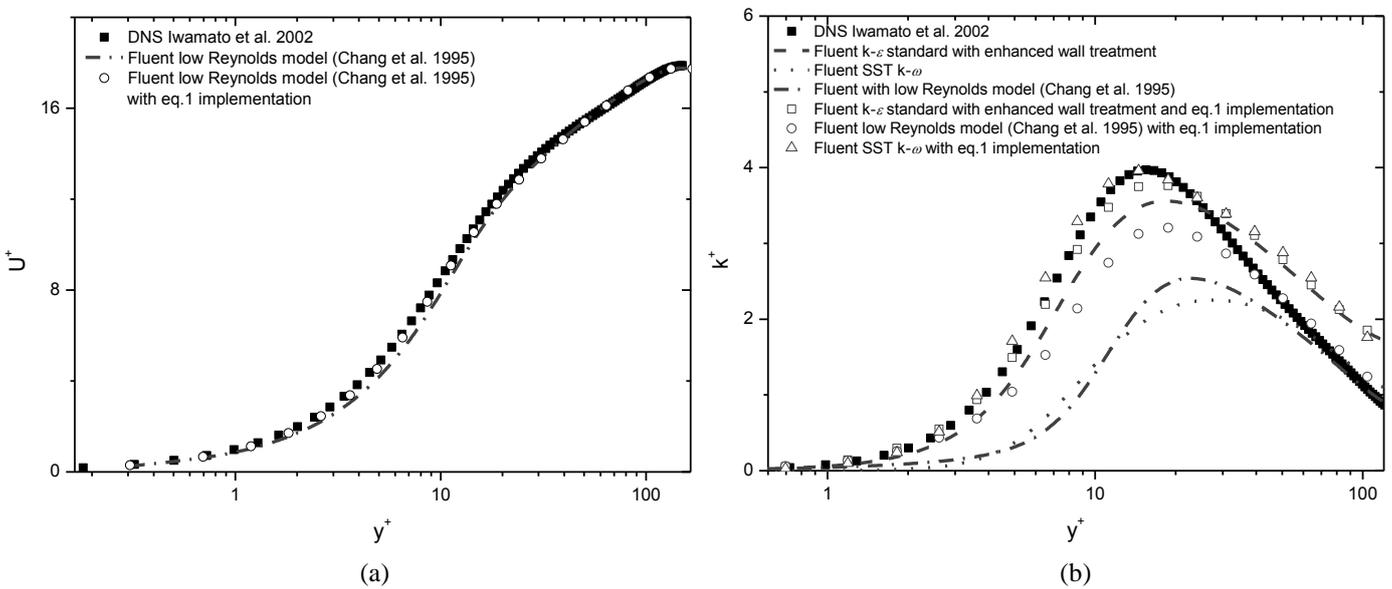

Figure 4 : Improved velocity (a) and turbulent kinetic energy (b) profiles with UDF implemented in Fluent

Figure 4 shows the improvement that can make the implementation of the function (1) in Fluent in the region $y+ \leq 30$ for the velocity and turbulent kinetic energy profiles.

**Anisotherm case.** Eq. (1) is designed to improve flow for an isothermal case, but it may also improve the flow in the anisothermal case because the error from the DNS is 3% (Fig. 5).

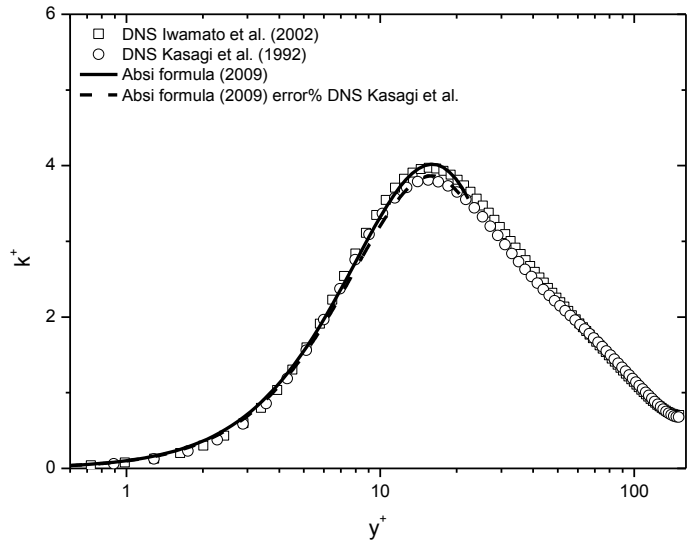

Figure 5 : Comparison of Eq. (1) to DNS $Re_\tau=150$

The three turbulence models: $k$-$\varepsilon$ standard with enhanced wall treatment , *SST k-ω* and low Reynolds model of Chang et *al*. predict in the same way the heat transfer in the region $y^+ \leq 10$. for the other values of $y^+$, the low Reynolds model of Chang et *al*. over-predicts DNS (Fig. 7.b ). The implementation of the UDF (Fig. 7b) leads to a decrease in the prediction of heat transfer. For low Reynolds model of Chang et *al.*, this decrease of the temperature profile makes the best prediction.

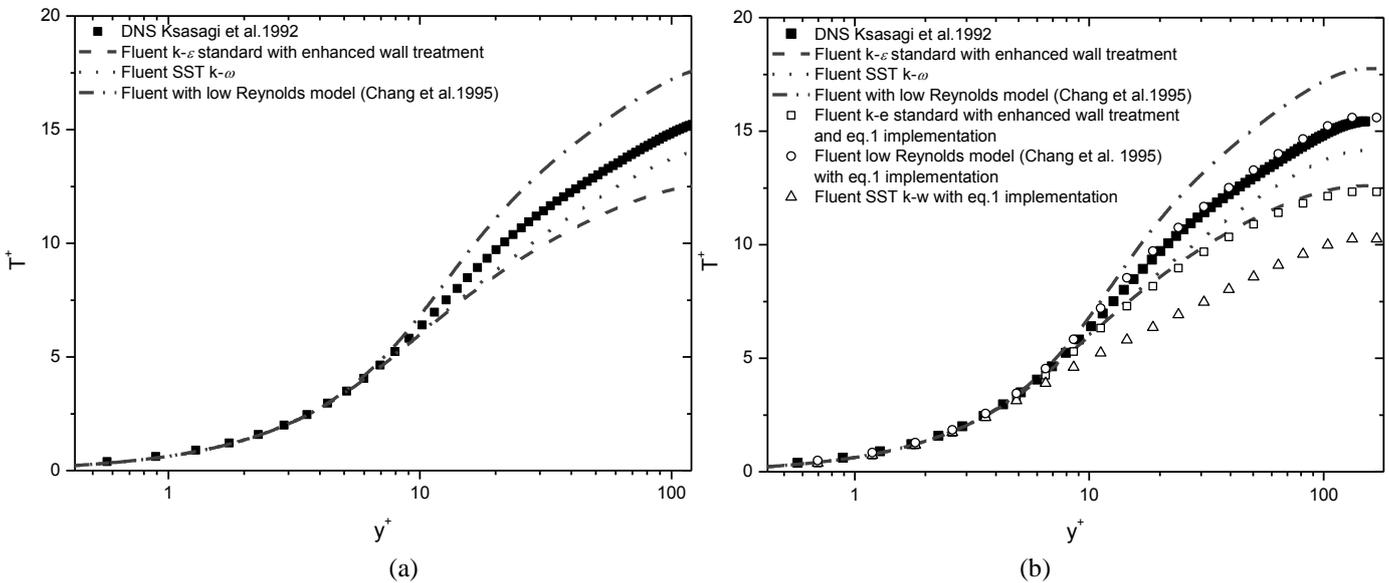

(a)     (b)

Figure 7 : Choice of turbulence model for heat transfer

**Conclusion**

Numerical simulations of isothermal and anisothermal channel flows are studied. A near-wall function for TKE is implemented in Fluent through a UDF to improve the prediction of velocity, turbulent kinetic energy and temperature profiles.

Three turbulent models: $k$-$\varepsilon$ standard with enhanced wall treatment, *SST k-ω* and low Reynolds model of Chang et *al.* were tested.

Results show that the implementation of this function in Fluent allows to improve the low Reynolds model of Chang *et al.* for velocity and temperature profiles.

**Acknowledgements**

The author would like to thank Prof. A. Baïri from IUT ville d'Avray for his hospitality in his laboratory and for the use of Fluent.

**References**

[1]     J. Kim and P. Moin, Turbulent Shear Flows 6, 85, (1987).

[2]     N. Kasagi, , Y. Tomita and A. Kuroda, Direct Numerical Simulation of Passive Scalar Field in a Turbulent Channel Flow, ASME J. Heat Transfer, 114(3), pp. 598–606, (1992).

[3]     H. Kawamura, H. Abe and Y. Matsuo, Int. J. Heat Fluid Flow, volume 20, pp 196–207 (1999).

[5]     I.Tiselj, R.Bergant, B.Mavko, I.Bajsic and G. Hetsroni, Journal of Heat Transfer, Vol. 123, pp 849- 857 (2001).

[6]     B.E. Launder, D.B, Spalding , Academic Press, London, (1972).

[7]     W.P. Jones, B. E. Launder , Int. J. Heat Mass Tran. 16: 1119-1130, (1973).

[8]     C. K. G. Lam, and K. Bremhorst, Journal of Fluids Engineering, Vol. 103, p. 456-460, (1981).

[9]     K. Y. Chien, AIAA Journal, Vol. 20, No. 1, pp. 33-38, (1982).

[10]    K. Abe, T. Kondoh and Y. Nagano , Int. J. Heat Mass Tran. 37: 139-151, (1994).

[11]    K. C. Chang, W. D. Hsieh, and C. S. Chen, Journal of Fluids Engineering, Vol. 117, September, p.17-423, (1995).

[12]    D. C. Wilcox , AIAA J. 33: 247-255, (1994).

[13]    F. R. Menter, NASA TM-103975. NASA-Ames research center, (1992).

[14]    Fluent Inc. 2005. Fluent 6.2 user's guide.

[15]    R. Absi, Analytical solutions for the modeled k-equation, ASME Journal of Applied Mechanics, 75(4), 1-4, (2008).

[16]    N. El Gharbi, R. Absi, A. Benzaoui, Int. J. CFD, accepted with corrections (2010).

[17]    R. Absi, A simple eddy viscosity formulation for turbulent boundary layers near smooth walls. Elsevier, C. R. Mecanique, 337, 158-165, (2009).

[18]    K. Iwamoto, Database of fully developed channel flow, THTLAB Internal Report No. ILR-0201, Dept. Mech. Eng., Univ. Tokyo, (2002).